\documentstyle[prl,floats,aps]{revtex}
\begin{document}
\draft
\begin{title}
{d-wave bipolaronic stripes and two energy scales in cuprates }
\end{title} 
\author{A.S. Alexandrov}
\address
{Department of Physics, Loughborough University, Loughborough LE11 
3TU, United Kingdom}

\maketitle
\begin{abstract}
There is strong experimental evidence for pairing of polaronic
 carriers in the normal state, two distinct energy scales, d-wave superconducting order parameter,
 and charge segregation in the form of stripes in several  cuprates.
 All   these remarkable phenomena  might be unified in the framework
 of the bipolaron theory  as
 a result of the formation of mobile  bipolarons in the normal state  and their
 Bose-Einstein condensation. Extending the BCS theory towards an
 intermediate and strong-coupling regime we show that  there are two
 energy scales in this regime, a temperature independent 
 incoherent gap  and a temperature dependent 
coherent gap   combining into one temperature
dependent global gap. 
The temperature dependence of the 
  gap and single particle (Giaver) tunnelling spectra  in cuprates are
  quantitatively described. A framework for understanding of two
 distinct energy scales observed  in Giaver tunnelling and Andreev  reflection
experiments is provided. We suggest that both d-wave  
 superconducting order parameter and striped charge distribution  result from the bipolaron
 (center-of-mass) energy band dispersion rather than from any
 particular interaction.
\end{abstract}
\narrowtext

\twocolumn

{\bf Introduction}
\vspace{0.5cm}

\noindent 
There is strong evidence for normal state gaps in high-$T_{c}$ 
cuprates from the uniform susceptibility \cite{jef,mul}, inelastic
neutron scattering and NMR
\cite{ros,nmr}, electron energy-loss  spectroscopy \cite{dem}, specific heat 
\cite{lor}, angle-resolved photoemission (ARPES) \cite{she,biaar}, 
tunnelling \cite{ren,brus2}, photoexcited quasiparticle relaxation
\cite{mic}, and  some 
kinetic measurements \cite{bat}. One view supported by ARPES  is that the 
gap reflects precursor superconducting correlations in the BCS-like state 
below some  characteristic temperature $T^{*}$ without long range phase 
coherence. Testing of this hypothesis with specific heat  \cite{lor} and 
tunnelling \cite{ren} data find that it cannot be sustained. In particular,
there is no sign that the gap 
closes at a given temperature $T^{*}$.  On the other hand 
the strong-coupling extension of the BCS theory based on the  
multi-polaron perturbation technique  firmly predicts the
transition to a charged Bose liquid in the crossover region  of  
  the BCS 
coupling constant $\lambda\simeq 1$ \cite{alex}. 
   (Bi)polaronic nature of carriers in  cuprates, supported by the 
   infrared  spectroscopy and quite a few other experiments\cite{tim}, in
   particular by the isotope effect on the 
   carrier mass \cite{muli}   provides a natural 
   microscopic explanation of the normal state gap  \cite{alem}.
   In the framework of the bipolaron theory the ground state of cuprates 
   is a charged 
Bose-liquid of intersite mobile  bipolarons where single polarons 
exist  as excitations with the energy $\Delta \sim T^*$ or larger
\cite{alemot}. 
A characteristic temperature $T^{*}$ of the normal phase is a 
crossover temperature of the order of $\Delta$ where the population of 
the  polaronic band becomes comparable with the bipolaron density.
Along this line the normal state kinetics of cuprates has  been explained 
\cite{alemot} and a theory of tunnelling \cite{aletun} and
angle-resoleved photoemission \cite{aleden} in cuprates has been developed 
describing essential spectral features of a single hole in these doped
Mott insulators.
 
However the  temperature and doping
dependence of the gap still remains a subject of controversy. Moreover, 
reflection experiments, in which an incoming electron from the normal
 side of a normal/superconducting contact is reflected as a hole along
 the same trajectory \cite{and0}, revealed a much smaller  gap edge than the
 bias
at the tunnelling conductance maxima in a few  underdoped cuprates
 \cite{yag}.  Recent intrinsic tunnelling measurements on a series of
 Bi '2212' single crystals \cite{pmul} showed distinctly different
 behaviour of the superconducting and normal state gaps with the
 magnetic field. Such coexistance of two distinct gaps in cuprates
 is not well understood \cite{deu2,pmul}.
There is also  a $d$-like  superconducting order parameter (changing sign when the
$CuO_{2}$ plane is rotated by $\pi/2$) in cuprates as observed in the  phase-sensitive
experiments \cite{pha}.
And, finally, there is
growing experimental evidence \cite{bia,tran} that stripes occur in
doped  cuprates. The $CuO_{2}$ plane of high-T$_{c}$ cuprates
has been found to be inhomogeneous on the microscopic scale,
suggesting that holes condense into large stripe domains \cite{bia2}.

Here I suggest  that two distinct energy scales, d-wave superconducting order parameter,
 and charge segregation in cuprates are manifestations of
 one and the same phenomenon, which is the normal state pairing of
 polaronic holes  in the doped charge-transfer Mott insulators.
The  main   assumption is that a short-range  attraction potential in
cuprates is large compared with  the (renormalised) polaron Fermi energy. Our  main point  is  independent of
 the microscopic nature of the attraction.  Real-space  pairs  might be
lattice and/or spin bipolarons \cite{alemot}, or any other preformed pairs.

\vspace{0.5cm}

{\bf 2. Normal and superconducting gap in cuprates}
\vspace{0.5cm}

\noindent 
Recently we have proposed  a toy  model
\cite{and}, which accounts for  two  energy scales. The model is a one-dimensional  Hamiltonian including 
kinetic energy of carriers in an effective mass ($m$) approximation and
a  local attraction potential, $V(x-x')=-U \delta(x-x')$ as
\begin{eqnarray}
 H &=& \sum_{s}\int dx \psi_s^{\dagger}(x)
 \left(-{1\over{2m}}{d^2\over{dx^2}} -\mu \right) \psi_s(x)\cr
&-& U \int dx \psi_\uparrow^{\dagger}(x)\psi_\downarrow^{\dagger}(x)\psi_\downarrow(x)\psi_\uparrow(x),
 \end{eqnarray}
 where $s= \uparrow,\downarrow$ is the 
 spin ($\hbar=k_{B}=1$). The first band to be doped in cuprates is the
 oxygen band inside the Hubbard gap as established in  polarised
 photoemission  \cite{aleden}. This band is almost one
 dimensional as discussed in Ref. \cite{aletun}, so that our
(quasi) one-dimensional approximation is a realistic starting point.
Solving a two-particle problem with the $\delta$-function potential
one obtains a bound state with the binding energy 
$2\Delta_{p}= {1\over{4}}mU^2$,
and with the radius of the bound state $r=2/mU$. We assume that this radius is less
than the inter-carrier distance in  cuprates, which puts a
constraint on the  doping level, $E_F < 2\Delta_p$, where $E_F$ is the polaron
Fermi energy.  Then  real-space  pairs are formed. If   three-dimensional corrections to the
energy spectrum of pairs are taken into account (see, for example,
Ref. \cite{alekab2}) 
 the ground state of the system is the Bose-Einstein condensate (BEC).  The
 chemical potential is pinned
 below the band edge by about  $\Delta_p$  both in the superconducting
 and normal state \cite{alemot}, so that the normal state
single-particle gap is $\Delta_p$.

Two distinct energy scales appear naturally if   we take into
account that in the superconducting state ($T<T_c$)
the single-particle excitations
 interact with the condensate via the same potential $U$. Applying the
 Bogoliubov approximation \cite{bog} we reduce the Hamiltonian, Eq.(1) to a
 quadratic form as
\begin{eqnarray}
 H &=& \sum_{s}\int dx \psi_s^{\dagger}(x)
 \left(-{1\over{2m}}{d^2\over{dx^2}} -\mu \right) \psi_s(x)\cr
&+& \int dx [\Delta_c \psi_\uparrow^{\dagger}(x)\psi_\downarrow^{\dagger}(x)+H.c.],
 \end{eqnarray}
where the coherent pairing potential,
$\Delta_c=-U\langle \psi_\downarrow(x)\psi_\uparrow(x)\rangle$,
is proportional to the square root of the condensate density,
$\Delta_c =constant \times  n_0(T)^{1/2}$.  The single-particle excitation energy
spectrum $E(k)$ is found using the Bogoliubov transformation as
\begin{equation}
E(k)= \left[(k^2/2m+\Delta_p)^2+\Delta_c^2\right]^{1/2}.
\end{equation}
 This spectrum is quite different
from the BCS quasiparticles because the chemical potential is negative
with respect to the bottom of the single-particle band,
$\mu=-\Delta_p$. The single particle gap, $\Delta$, defined as  the minimum of
$E(k)$, is given  by
\begin{equation}
\Delta= \left[\Delta_p^2+\Delta_c^2 \right]^{1/2}.
\end{equation}
It varies with temperature from $\Delta(0)=
\left[\Delta_p^2+\Delta_c(0)^2 \right]^{1/2}$ at zero temperature down to
the temperature independent $\Delta_p$  above $T_c$. The condensate
density depends on temperature as $1-(T/T_c)^{d/2}$ in the ideal
 three ($d=3$) and (quasi) two-dimensional ($d=2$) Bose-gas.  In the three-dimensional $charged$ Bose-gas it has an
exponential temperature dependence at low temperatures due to a plasma
gap  in the Bogoliubov  excitation spectrum \cite{fol}, which might
be highly anisotropic in cuprates \cite{alemot}. Near T$_c$ 
one can expect a power law dependence, $n_0(T) \propto 1- (T/T_c)^n$
with $n>d/2$  because the condensate plasmon \cite{fol}
depends on temperature.
As shown in Ref. \cite{and} the theoretical temperature dependence, Eq.(4) describes well  the pioneering experimental observation of
the anomalous gap in $YBa_2Cu_3O_{7-\delta}$ in the
electron-energy-loss spectra by Demuth $et$ $al$ \cite{dem}, with
$\Delta_c(T)^2=\Delta_c(0)^2 \times[1-(T/T_c)^n]$ below $T_c$ and 
 zero above $T_c$, and $n=4$.

\vspace{0.5cm}

{\bf 3. Giaver tunnelling  and Andreev reflection} 
\vspace{0.5cm}

\noindent   
 The normal metal-superconductor (SIN)  tunnelling conductance  via a dielectric contact, $dI/dV$ is proportional
 to the density of states, $\rho(E)$  of the spectrum 
 Eq.(3).  Taking also into account the scattering of single-particle excitations
 by a random potential, thermal lattice and
 spin fluctuations as described in Ref. \cite{aletun} one  finds at
 $T=0$ 
\begin{equation}
dI/dV \propto 
[\rho\left({2eV-2\Delta\over{\epsilon}}\right)+
A\rho\left({-2eV-2\Delta\over{\epsilon}}\right)],
\end{equation}
with
\begin{equation}
\rho(\xi)={4\over{\pi^{2}}}\times{Ai(-2\xi)
Ai'(-2\xi)+
Bi(-2\xi)Bi'(-2\xi)\over{
[Ai(-2\xi)^{2}
+Bi(-2\xi)^{2}]^{2}}},
\end{equation}
$A$ is the asymmetry coefficient\cite{aletun}, $Ai(x), Bi(x)$ the Airy functions, and $\epsilon$ is the
scattering rate. 
We compare the  conductance, Eq.(5) with  one of the best
 STM
 spectra measured in $Ni$-substituted $Bi_2Sr_2CaCu_2O_{8+x}$ single
crystals by Hancottee $et$ $al$\cite{brus1}, Fig.1a. This experiment showed anomalously
large $2\Delta/T_c > 12$.

The theoretical conductance, Eq.(5) describes well 
 the 
anomalous $gap/T_{c}$ ratio, injection/emission assymmetry,
zero-bias conductance at zero temperature,  and the spectral shape inside and 
outside the gap region. There is no doubt that the gap, Fig.1 is
s-like. The  conductance,
Eq.(5) fits also well the  conductance curve
obtained on  'pure' Bi2212 single crystals \cite{brus1}, while
a simple d-wave BCS denstity of states cannot describe the excess
spectral weight in  the peaks and the shape  of the conductance outside the
gap region. We notice that the scattering rate,
$\epsilon$, derived from the fit, is apparently smaller in the 'pure' sample than in 
$Ni$- substituted one, as expected.

The same toy model provides  a simple theory of  tunnelling into bosonic
 (bipolaronic) superconductor in the metallic (no-barrier)
regime \cite{and}. As in the canonical BCS approach applied to the
 Andreev
 tunnelling  by Blonder, Tinkham and
Klapwijk \cite{btk}, the incoming electron produces
only outgoing particles in the superconductor ($x>l$), allowing for a
 reflected electron and (Andreev) hole in the normal
metal ($x<0$). There is also a buffer layer of the thickness $l$ at the normal
metal-superconductor boundary ( $x=0$), where the chemical potential 
with respect to the bottom of the conduction band changes gradually
from a positive large value $\mu$ in the metal to a negative 
value $-\Delta_p$ in the bosonic superconductor. We approximate this buffer layer  by 
 a layer with a constant chemical potential $\mu_b$ ($-\Delta_p<\mu_{b}<\mu$) and with the
 same strength of the pairing potential $\Delta_c$ as in the bulk
 superconductor.  The Bogoliubov-de Gennes equations may be written
as usual  \cite{btk}, with the only difference that the chemical
 potential with respect to the bottom of the band is a function of the
 coordinate $x$,
\begin{eqnarray}
\left 
(\matrix{-(1/2m) d^2/dx^2 -\mu(x)& \Delta_c \cr \Delta_c &(1/2m)
  d^2/dx^2 +\mu(x) }\right) \psi(x) \cr
= E\psi(x) 
\end{eqnarray}
  Thus the two-componet wave function in the normal
 metal is given by  
\begin{equation}
\psi_n(x<0)=\left 
(\matrix{1 \cr 0 }\right)e^{iq^+x}+b\left 
(\matrix{1 \cr 0 }\right)e^{-iq^+x}+a\left 
(\matrix{0\cr 1 }\right)e^{-iq^{-}x},
\end{equation}
while in the buffer layer it has the form
\begin{eqnarray}
\psi_b(0<x<l)&=&\alpha \left 
(\matrix{1 \cr {\Delta_c\over{E+\xi}} }\right)e^{ip^+x}+\beta \left 
(\matrix{1 \cr  {\Delta_c\over{E-\xi}}}\right)e^{-ip^{-}x}\cr
&+&\gamma\left 
(\matrix{1\cr  {\Delta_c\over{E+\xi}} }\right)e^{-ip^+x}+\delta \left 
(\matrix{1 \cr  {\Delta_c\over{E-\xi}}}\right)e^{ip^{-}x},
\end{eqnarray}
where  the momenta associated with the energy $E$ are
$q^{\pm}=[2m(\mu \pm E )]^{1/2}$
and
$p^{\pm}=[2m(\mu_b \pm \xi )]^{1/2}$
with $\xi= (E^2-\Delta_c^2)^{1/2}$. The well-behaved solution in the
superconductor with  negative chemical potential is given by
\begin{equation}
\psi_s(x>l)=c \left 
(\matrix{1 \cr  {\Delta_c\over{E+\xi}} }\right)e^{ik^+x}+d \left 
(\matrix{1 \cr {\Delta_c\over{E-\xi}}   }\right)e^{ik^{-}x},
\end{equation}
where  the momenta associated with the energy $E$ are
$k^{\pm}=[2m(-\Delta_p \pm \xi )]^{1/2}$.
The coefficients $a,b,c,d, \alpha,\beta,\gamma,\delta$ are determined
from the boundary conditions, which are continuity of $\psi(x)$ and
its first
derivative at $x=0$ and  $x=l$. Applying the boundary conditions,
and carrying out an algebraic reduction, we find
\begin{equation}
a= 2\Delta_c q^+(p^+f^-g^+ -p^-f^+g^-)/D,
\end{equation}
\begin{eqnarray}
b&=&-1 +2q^+[(E+\xi)f^+(q^-f^--p^-g^-)/D \cr
&-& 2q^+(E-\xi)f^-(q^-f^+-p^+g^+)]/D,
\end{eqnarray}
with
\begin{eqnarray}
D&=&(E+\xi)(q^+f^++p^+g^+)(q^-f^--p^-g^-)\cr
&-&(E-\xi)(q^+f^-+p^-g^-)(q^-f^+-p^+g^+),
\end{eqnarray}
and $f^\pm= p^\pm \cos(p^\pm l)-ik^\pm \sin(p^\pm l)$, $g^\pm=k^\pm
\cos(p^\pm l) -i p^\pm \sin(p^\pm l)$.

The transmisson coefficient for electrical current,
$1+|a|^2-|b|^2$ is shown in Fig.2 
for different values of $l$ when the coherent gap $\Delta_c$ is
smaller than  the pair-breaking gap $\Delta_p$.  We find two
distinct energy scales, one is $\Delta_c$ in the subgap region due to 
electron-hole  reflection and the other one is $\Delta$, which is the
single-particle band edge.  We notice that the transmission has no
subgap structure if the buffer layer is absent ($l=0$) in both cases. In the
extreme case of a wide buffer layer, $l>> (2m\Delta_p)^{-1/2}$, Fig.2.
 there are some oscillations of the
transmission due to the bound states inside the buffer layer. It was
shown in Ref. \cite{alekab3} that the pair-breaking gap $\Delta_p$  is inverse
proportional to the doping level. On the other hand, the coherent gap $\Delta_c$ 
scales with the condensate density, and therefore with the critical
temperature, determined as the Bose-Einstein condensation temperature
of strongly anisotropic 3D bosons \cite{alekab2}.
Therefore we expect that $\Delta_p>>\Delta_c$ in the underdoped
cuprates, Fig.2. Thus the model accounts for the two different gaps experimentally
observed in  Giaver tunnelling and electron-hole reflection in the
underdoped cuprates \cite{deu2}.

One of the important experimental findings is the zero-bias
conductance peak (ZBCP) observed in a number of in-plane tunnel
junctions 
\cite{ZBC}.  Within our approach, ZBCP-like
feature appears  in the junctions with
$\Delta_p > \Delta_c$ and $l>>1$ (see thin dashed line in Fig.2).
 Overdoped cuprates might
 be in the BCS or intermediate regime because the increasing number of
 holes leads to an increase of their Fermi energy and, at the same
 time, to  a decrease of $\Delta_p$ due to screening
 \cite{alekab3}. Then,  the condition for BEC, $E_F<2\Delta_p$ 
 no longer holds, and the d-wave BCS approach \cite{the} might account  for 
 ZBCP as well. Such evolution from the BEC regime  in underdoped cuprates
 to the  BCS-like
 regime
 in the overdoped samples has been  found in the   conserving T-matrix
 approximation of the two-dimensional attractive Hubbard
 model \cite{kor}.  The binding energy of real-space pairs
monotonically decreased with increasing density,  and vanished at a
critical density of holes 
$n_{cr} = 0.19$ (per cell).

{\vspace{0.5cm}

{\bf 4. D-wave Bose-Einstein condensate}

\vspace{0.5cm}

\noindent 
To account for the symmetry of
the order parameter and  the 
orientation dependence of ZBCP our simple
one-dimensional continuos  model  can be readily
extended to two and three dimensions, discrete lattices,  and nonlocal
pair potential. 
Than, quite generally,  the  symmetry of the single-particle
spectrum, Eq.(3) and 
the symmetry of the Bose-Einstein condensate
wave-function (i.e. of the order parameter)  are not necessary the
same in the bosonic superconductor \cite{aletun}. The single-particle spectrum is dominated by the incoherent
(s-wave) gap, $\Delta_p$, in agreement with the c-axis tunnelling data,
Fig.1. On the other hand, the
symmetry of the Bose-Einstein condensate depends on the  bosonic pair (center
of mass) energy band dispersion.

Different
 bipolaron configurations can be found  with computer
simulation techniques based on the minimization of the ground state
energy of an ionic lattice with two holes. The intersite pairing of
the in-plane oxygen hole  with the  $apex$ one is
energetically favorable in the layered perovskite structures as
 established by
Catlow $et$ $al$ \cite{cat}.
This apex or peroxy-like bipolaron can tunnel from one cell to another  via a
direct $single$ $polaron$ tunnelling from one apex oxygen to its apex
neighbor.
 The bipolaron band structure has been derived in Ref.\cite{aler}
as
\begin{equation}
E^{x,y}_{\bf k}=tcos( k_{x,y})-t'cos( k_{y,x}).
\end{equation}
Here the in-plane lattice constant is taken as $a=1$,
 $t$ is twice the bipolaron
 hopping integral between $p$ orbitals of the same symmetry
elongated in the direction
 of the hopping ($pp\sigma$)
 and $-t'$ is twice the hopping integral in the perpendicular
 direction ($pp\pi$). These hopping integrals are proportional to the
 single-particle hopping integrals between $apex$ oxygen ions.   The  bipolaron energy spectrum in  the tight
binding approximation consists of two bands $E^{x,y}$ formed by the
overlap of
$p_{x}$  and  $p_{y}$ $apex$ $polaron$ orbitals, respectively.
The energy band minima are found at the Brillouin zone
 boundary,  ${\bf
 k}=(\pm \pi,0)$ and  ${\bf k}=(0, \pm \pi)$ rather than at the  $\Gamma$
point
 owing to the opposite sign of the $pp\sigma$ and $pp\pi$
 hopping integrals.  
Neither band is 
invariant under crystal symmetry but the degenerate doublet is
an irreducible representation; under a $\pi/2$ rotation the $x$ band
transforms into $y$ and vice versa.

If the bipolaron  density  is low, the
 bipolaron Hamiltonian can be mapped onto the charged Bose-gas
 \cite{alemot}. Charged bosons are condensed below $T_{c}$ into the
 states of the Brillouin zone with the lowest energy, which are ${\bf
 k}=(\pm \pi,0)$ and  ${\bf k}=(0, \pm \pi)$ for the $x$ and $y$
 bipolarons, respectively. These four states are degenerate, so the
 condensate field-operator $\Psi({\bf m})$  in the real
(site) space ${\bf m}= (m_{x},m_{y})$  is given
 by
 \begin{equation}
\hat{ \Psi}_{\pm}({\bf m}) = N^{-1/2}\sum_{{\bf k}=(\pm \pi,0),(0,\pm \pi)} b_{\bf
k} exp(-i{\bf k \cdot m}).
 \end{equation}
 where $N$ is the number of cells in the crystal, and  $b_{\bf k}$ is the
 bipolaron
(boson) annihilation operator in
 ${\bf
 k}$ space. This is   a $c$-number below T$_c$, so that the condensate
 wave function, which is the superconducting order parameter, is
 given by
\begin{equation}
\Psi_{\pm}({\bf m})= n_c^{1/2} \left[ \cos (\pi m_{x}) \pm \cos
 (\pi m_{y})\right],
\end{equation}
where $n_c$ is the
 number of bosons per cell in the condensate.
 Other combinations of the four degenerate states do not respect
 time-reversal and (or) parity symmetry.
The two solutions, Eq.(16),  are physically identical being
 related by:
 $\Psi_{+}(m_{x},m_{y})=\Psi_{-}(m_{x},m_{y}+1)$. They have
 $d$-wave  symmetry changing  sign when the
$CuO_{2}$ plane is rotated by $\pi/2$ around $(0,0)$  for $\Psi_{-}$ or
around (0,1) for $\Psi_{+}$.
  The
$d$-wave symmetry is entirely due to the bipolaron energy dispersion
with four minima at ${\bf k \neq 0}$.

\vspace{0.5cm}

{\bf 5. Bipolaronic stripes}
\vspace{0.5cm}

\noindent 
The
antiferromagnetic interaction are  thought to give rise 
to charge segregation (stripes) in cuprates\cite{zaa,emkiv}. At the
same time
the electron-phonon interaction is strong in ionic cuprates and manganites
\cite{alemot,alebra}. While for wide band  polar semiconductors
and polymers  the charge segregation (strings)
were discussed some  time ago \cite{gri1}, a  role  of the  long-range  Fr\"ohlich
electron-phonon  interaction in the charge segregation in $narrow$
band ionic insulators like cuprates has been addressed only recently
\cite{kus,alekabst} with the  opposite conclusions about  
existance of strings.

We  have proved \cite{alekabst} that the Fr\"ohlich
interaction combined with the direct Coulomb repulsion  does not lead
to  charge  segregation like  strings in doped
narrow-band insulators, both in the nonadiabatic and adiabatic regimes.
The polarisation  potential energy of $M$ fermions trapped in a string  of the length $N$,
\begin{equation}
U=-{e^{2}\over{\kappa}} M^{2} I_{N}, 
\end{equation}
appears to be lower in value than the
repulsive Coulomb energy because  $\kappa=\epsilon_{0}\epsilon_{\infty}/(\epsilon_{0}-\epsilon_{\infty})$
is always larger  than the high-frequency dielectric constant $\epsilon_{\infty}$.
  Here $\epsilon_{0}$ is the static dielectric constant, and  the integral $I_{N}$ is given by
\begin{eqnarray}
 I_N&=& {\pi \over{(2\pi)^3}} \int _{-\pi}^\pi dx \int _{-\pi}^\pi dy  \int
     _{-\pi}^\pi dz {\sin(Nx/2)^2\over{N^2 \sin(x/2)^2}}\cr
&\times&( 3-\cos x -\cos
     y -\cos z)^{-1}.
\end{eqnarray}
It has the following asymptotics at large $N$
\begin{equation}
I_{N}= {1.31+\ln N\over{N}}.
\end{equation}
The Coulomb energy of spinless one-dimensional fermions comprising both
 Hartree and exchange terms is \cite{ref} 
\begin{equation}
  E_C={e^2M(M-1)\over{N\epsilon_{\infty}}} [0.916+\ln M],
\end{equation}
so that  the polarisation and Coulomb  energy  per particle  becomes 
\begin{equation}
U/M={e^2M\over{N \epsilon_{\infty}}}[0.916+\ln M - \alpha(1.31+\ln N)],
\end{equation}
where $\alpha= 1-\epsilon_{\infty}/\epsilon_0 <1$. Minimising this
energy with respect to the length of the string $N$ we find
\begin{equation}
N= M^{1/\alpha} \exp (-0.31+0.916/ \alpha), 
\end{equation}
and 
\begin{equation}
(U/M)_{min}= -{e^2\over{\kappa}}
M^{1-1/\alpha} \exp(0.31-0.916 /\alpha).
\end{equation}
One can see that the potential energy per particle increases
 with the number of particles.  Hence,  the
energy of $M$ well separated polarons is lower than the energy of  polarons trapped
in a string no matter correlated or not.  The opposite conclusion
 derived in   
Ref. \cite{kus} originates in an erroneous  approximation of the integral
 $I_N \propto N^{0.15}/N$. The correct asymptotic result is 
 $I_N = \ln(N)/N$. 
 
In prinicple, the situation might be different
if the antiferromagnetic interaction of doped holes with the parent Mott
insulator is strong enough \cite{zaa,emkiv}. Due to the  long-range
 nature of the Coulomb repulsion the length of a single stripe should
 be finite (see, also Ref.\cite{bia,kus2}). Including the Fr\"ohlich, Eq.(17)
 and Coulomb Eq.(20) contributions one can readily   estimate  its
 length
 for any type of the short-range $intersite$  attraction, $E_{att}$ as \cite{alekabst}
\begin{equation}
N = \exp \left(\frac {\epsilon_{0}aE_{att} \delta \omega }{e^{2}\omega} 
-2.31 \right ),
\end{equation}
where $\omega$ is the characteristic frequency of bosonic excitations
(like magnons) responsible for the short-range attraction,
$\delta \omega$ its maximum dispersion, and $a$ is the lattice
constant.  Then, with typical values of $\epsilon_{0}=30$, $a=3.8 \AA$ and
$E_{att}=2J=0.3$ eV of the $t-J$ model one can hardly expect any charge segregation due
to antiferromagnetic spin fluctuations as well, because  $N < 2$  according to Eq.(24).
Hence  it is more probable that the  lattice and spin polarons in cuprates
remain in a liquid state
in the relevant region of the parameters. 

The Fr\"ohlich interaction is not the only electron-phonon
interaction in  ionic solids. As
discussed in Ref. \cite{alemot}, any short range electron-phonon
interaction, like, for example, the  deformation potential and/or Jahn-Teller distortion can overcome
 the residual weak repulsion 
of Fr\"ohlich  polarons to form small
bipolarons. At large distances small bipolarons weakly  repel
each other due to the long-range Coulomb interaction, strongly
suppressed by optical phonons.  Hence, they  form a liquid state
 in the relevant region of their densities and masses \cite{ale}, which is also
 confirmed by recent study \cite{bon}  of
the Holstein-Hubbard model.  The ground state of the 1D Holstein-Hubbard model is a
liquid of intersite bipolarons with a significantly reduced mass 
\cite{bon}. The
bound states of three or more polarons are not stable. 

If (bi)polaronic
carriers in cuprates are in the liquid state, one
can pose a key question how stripes can be seen at all.
Here I suggest that the bipolaron liquid might be striped owing to the
bipolaron energy band dispersion, Eq.(14), rather than to any particular
interaction. If  bipolaron  bands have their minima
at finite {\bf k} (including the Brillouin zone boundaries), then
the  condensate wave function
is modulated in the   real space Eq.(16).  As a result, the  hole density, which is
about  twice of  the  condensate density at low temperatures,
is striped, 
with the characteristic period of  stripes
determined by the inverse band-minima wave vectors. If disorder is strong  enough, the condensate might be
localised within the  wide and shallow random potential wells, so bipolaronic stripes
can also exist in the insulating (due to disorder) oxides.  If the
condensate is striped, one can  
expect rather different values of the  coherent gap,
$\Delta_c$,  for single-particle excitations  tunnelling along (110) and
(100), and,  hence the
different Andreev reflection spectra. Our
interpretation of stripes is consistent with recent inelastic neutron
scattering studies   of YBa$_{2}$Cu$_{3}$O$_{7-\delta}$ \cite{bou,bou2}, where
the incommensurate peaks in neutron scattering have been  only observed in the
$superconducting$ state. The vanishing at T$_{c}$ of the
incommensurate peaks can be actually anticipated over a wide part of
the phase diagram and in  other neutron  data \cite{moo}. It is,
of course, inconsistent with the usual stripe picture where a
characteristic distance needs to be observed in the normal state as
well. In our picture with  the d-wave $striped$  Bose-Einstein condensate 
the incommensurate neutron  peaks disappear at T$_{c}$ or slightly above
T$_{c}$ (due to superconducting fluctuations), as observed. 

To conclude I suggest that two distinct energy scales, d-wave superconducting
order parameter and stripes in underdoped cuprates are  
manifestations of  mobile bipolarons. The normal state gap in the
charge channel is
 half of the bipolaron binding energy. The superconducting gap
comprises additional  coherent part owing to the interaction of single
holes with the Bose-Einstein condensate. d-wave  superconducting
order parameter and
stripes are intimately connected as  the result of the bipolaron energy band dispersion
with the minima at finite wave vectors of the center-of-mass Brillouin zone. 

I  greatly appreciate enlightening  discussions with Antonio
Bianconi and   Robert  Laughlin.

\centerline{{\bf Figure Captures}}

Fig.1. Theoretical tunnelling conductance, Eq.(5)  (line) compared with the  STM 
conductance 
in  Ni-substituted $Bi_2Sr_2CaCu_2O_{8+x}$ \cite{brus1} (dots)
with $2\Delta = 90$ meV,
$A = 1.05$, $\epsilon= 40$ meV.

Fig.2. Transmission versus voltage (measured in units of  $\Delta_p/e$)
for $\Delta_c=0.2\Delta_p$, $\mu=10 \Delta_p$, $\mu_b=2 \Delta_p$ and  
$l=0$ (thick  line), $l=1$ (thick dashed line), $l=4$ (thin line), and
$l=8$ (thin dashed line) (in units of $1/(2m \Delta_p)^{1/2}$).


\begin{references}
\bibitem{jef}
D.C.  Johnston, Phys. Rev. Lett. {\bf 62} 957 (1989)
\bibitem{mul}
K.A. M\"uller $et$ $al$, J.Phys.: Condens. Matter {\bf 10}, L291 
(1998).
\bibitem{ros}
J. Rossat-Mignot $et$ $al$, Physica (Amsterdam) B {\bf 18--181}, 383
(1992).
\bibitem{nmr}
For a comprehensive review see C. Berthier $et$ $al$, J. Phys. I
France {\bf 6}, 2205 (1996).
\bibitem{dem}
J.E. Demuth $et$ $al$, Phys. Rev. Lett. {\bf 64}, 603 (1990).
\bibitem{lor}
J.W. Loram $et$ $al$,
 J. of Superconductivity {\bf 7}, 243 (1994).
\bibitem{she}
Z.-X. Shen and J.R. Schrieffer, Phys. Rev. Lett. {\bf 78}, 1771
(1997) and references therein.
\bibitem{biaar}
N.L. Saini $et$ $al$, Phys. Rev. Lett. {\bf 79}, 3467 (1997).
\bibitem{ren}
Ch. Renner $et$ $al$, Phys. Rev. Lett. {\bf 80}, 149 (1998).
\bibitem{brus2}
A. Mourachkine, Europhys. Lett. {\bf 49}, 86 (2000).
\bibitem{mic}
V.V. Kabanov, J. Demsar, B. Podobnik, and D. Mihailovic, Phys. Rev. B
{\bf 59}, 1497 (1999).
\bibitem{bat}
B. Batlogg $et$ $al$, Physica C (Amsterdam) {\bf 135-140}, 130 (1994).
\bibitem{alex}
A.S. Alexandrov,  Zh.Fiz.Khim. ${\bf 57}$, 273 (1983)
 (Russ.J.Phys.Chem.${\bf 57}$, 167 (1983)).
\bibitem{tim}  D.B. Tanner and T. Timusk, in `Physical Properties of 
High-Temperature  
Superconductors III', ed. D.M.  Ginsberg, World 
Scientific, Singapore (1992);
E.K.H. Salje, A.S. Alexandrov, and W.Y. Liang (eds),  `Polarons and 
Bipolarons in High-$T_{c}$  
Superconductors and Related Materials', 
Cambridge University Press, Cambridge (1995), 
 J.T. Devreese, in
Encyclopedia of Applied Physics, vol. 14, p. 383, VCH Publishers (1996).
\bibitem{muli}
G. Zhao $et$ $al$,  Nature ${\bf 
 385}$, 236 (1997).
\bibitem{alem}
A.S. Alexandrov, Physica C (Amsterdam) ${\bf 182}$, 327 (1991).
\bibitem{alemot}
A.S. Alexandrov and N.F. Mott, Rep. Prog. Phys. ${\bf 57}$ 1197 
(1994);
{\it Polarons and Bipolarons}, 
World-Scientific (1995). 
\bibitem{aletun}
A.S. Alexandrov, Physica C (Amsterdam) {\bf 305}, 46 (1998).
\bibitem{aleden}
A.S. Alexandrov and C.J. Dent, Phys. Rev. B{\bf
  60}, 15414 (1999).
\bibitem{and0}
A.F. Andreev, Zh. Eksp. Teor. Fiz. {\bf 46}, 1823 (1964) [Sov.
Phys. JETP {\bf 19}, 1228 (1964)].
\bibitem{yag}
Y. Yagil $et$ $al$, Physica C (Amsterdam) {\bf 250}, 59 (1995).
\bibitem{pmul}
P. M\"uller $et$ $al$, unpublished.
\bibitem{deu2}
G. Deutscher, Nature {\bf 397}, 410 (1999).
\bibitem{pha}
 C.C. Tsuei $et$ $al$,
Science {\bf 272}, 329 (1996) and references therein.
\bibitem {bia}
A. Bianconi, J. Phys. IV France, {\bf 9}, 325 (1999) and references
therein.
\bibitem{tran}
J.M. Tranquada $et$ $al$, Nature {\bf 375}, 561 (1996).
\bibitem{bia2}
A. Bianconi $et$ $al$, Physica C {\bf 296}, 269 (1998).
\bibitem{and}
A.C. Alexandrov and A.F. Andreev, cond-mat/0005315, to be published.
\bibitem{alekab2}
A.S. Alexandrov and V.V. Kabanov, Phys. Rev. B {\bf 59}, 13628 (1999).
\bibitem{bog}
N. Bogoliubov, J.Phys. USSR {\bf 11}, 23-32(1947).
\bibitem{fol}
L.L. Foldy, Phys.Rev. {\bf 124}, 649(1961).
\bibitem{brus1}
H. Hancotte $et$ $al$, Phys. Rev. B{\bf 55}, R3410 (1997).
\bibitem{btk}
G.E. Blonder, M. Tinkham, and T.M. Klapwijk, Phys. Rev. B{\bf 25},
4515 (1982).
\bibitem{ZBC}
 L. Alff $et$ $al$, Phys. Rev. B{\bf 55}, R14757 (1997); M. Covington
$et$ $al$, Phys. Rev. Lett. {\bf 79}, 277 (1997); S. Sinha and K.W. Ng,
Phys. Rev. Lett {\bf 80}, 1296 (1998), W. Wang $et$ $al$, Phys. Rev. B
{\bf 60}, 4272 (1999) and references therein.
\bibitem{alekab3}
A.S. Alexandrov, V.V. Kabanov and N.F. Mott, Phys. Rev. Lett. ${\bf 
77}$, 4796 (1996).
\bibitem{the}
Y. Tanaka and S. Kashiwaya, Phys. Rev. Lett {\bf 74}, 3451 (1995).
\bibitem{kor}
B.Kyung, E.G. Klepfish and P.E.  Kornilovitch,
Phys. rev. Lett. {\bf 80}, 3109 (1998). 
\bibitem{cat}
C.R.A. Catlow, M.S. Islam and X. Zhang, J. Phys.: Condens. Matter 
${\bf 10}$, L49 (1998).
\bibitem{aler}
A.S. Alexandrov, Phys. Rev. B${\bf 53}$, 2863 (1996).
\bibitem{zaa}
J. Zaanen and O. Gunnarsson, Phys. Rev. B {\bf 40}, 7391 (1989).
\bibitem{emkiv}
V. J. Emery $et$ $al$, Phys. Rev. B {\bf 56}, 6120 (1997) and
references therein.
\bibitem{alebra}
A.S. Alexandrov and A.M. Bratkovsky, J. Phys.: Condens. Matter {\bf
  11}, L531 (1999).
\bibitem{gri1}
L.N. Grigorov, Makromol. Chem., Macromol. Symp. {\bf 37}, 159 (1990).
\bibitem{kus}
F.V. Kusmartsev, Phys. Rev. Lett., {\bf 84}, 530 (2000).
\bibitem{alekabst}
A.S. Alexandrov and V.V. Kabanov, cond-mat/0005419, to be published.
\bibitem{ref} 
This expression differs  numerically from
  Ref. \cite{kus2}.
\bibitem{kus2}
F.V. Kusmartsev, J. Phys. IV France, {\bf 9}, 321 (1999).
\bibitem{ale}
A.S. Alexandrov, Phys. Rev. B {\bf 61}, 12315 (2000).
\bibitem{bon}
J. Bonca $et$ $al$, Phys. Rev. Lett., {\bf 84}, (2000). 
\bibitem{bou}
P. Bourges $et$ $al$, Science {\bf 288}, 1234 (2000).
\bibitem{bou2}
P. Bourges $et$ $al$, cond-mat/0006085.
\bibitem{moo}
H.A. Mook $et$ $al$, Nature {\bf 395}, 580 (1998); M. Arai $et$ $al$,
Phys. Rev. Lett. {\bf 83}, 608 (1999); P. Dai $et$ $al$,
Phys. Rev. Lett. {\bf 80}, 1738 (1998). 










\end{references}
\end{document}